\documentclass{emulateapj}
\usepackage{graphicx}
\usepackage{subfigure}
\usepackage{epsfig}
\usepackage{times}
\usepackage{natbib}
\usepackage{amsfonts}
\usepackage{amsmath}
\usepackage{amsbsy}

\begin{document}
\title{Observable QPOs produced by steep pulse profiles in Magnetar Flares}
\author{C. R. D'Angelo and A. L. Watts}
\affil{Instituut Anton Pannekoek, University of Amsterdam, Amsterdam  1098 XH,
  The Netherlands}
\email{c.r.dangelo@uva.nl}

\begin{abstract}
  Strong quasi-periodic oscillations (QPOs) in the tails of the giant
  gamma- ray flares seen in SGR 1806--20 and SGR 1900+14 are thought
  to be produced by starquakes in the flaring magnetar. However, the
  large fractional amplitudes (up to $\sim$20\%) observed are
  difficult to reconcile with predicted amplitudes of starquakes. Here
  we demonstrate that the steeply pulsed emission profile in the tail
  of the giant flare can enhance the observed amplitude of the
  underlying oscillation, analogously to a beam of light oscillating
  in and out of the line of sight. This mechanism will also broaden
  the feature in the power spectrum and introduce power at harmonics
  of the oscillation. The observed strength of the oscillation depends
  on the amplitude of the underlying starquake, the orientation and
  location of the emission on the surface of the star, and the
  gradient of the light curve profile. While the amplification of the
  signal can be significant, we demonstrate that, even with
  uncertainties in the emission geometry, this effect is not
  sufficient to produce the observed QPOs. This result excludes the
  direct observation of a starquake and suggests that the observed
  variations come from modulations in the intensity of the emission.
\end{abstract} 

\section{Introduction}
\label{sec:introduction}

Giant gamma-ray flares are thought to be produced by a large-scale
rearrangement of the super-strong magnetic field of a magnetar
\citep{1995MNRAS.275..255T,2001ApJ...561..980T}. The large amount of
energy released during a flare ($\sim$10$^{46}$erg) likely triggers
large-scale vibrations and quakes in the star's crust
\citep{1998ApJ...498L..45D}. The observation of strong, long-lived
quasi-periodic oscillations (QPOs) in the decay tails of the giant
flares of SGR 1900+14 and SGR 1806--20 has been interpreted as
evidence of such starquakes
(\citealt{2005ApJ...628L..53I,2005ApJ...632L.111S,2006ApJ...637L.117W,2006ApJ...653..593S};
hereafter SW06), and much theoretical work has subsequently focused on
the physics of starquakes themselves
(e.g.,\citealt{2006MNRAS.368L..35L,2006MNRAS.371L..74G,2011MNRAS.410L..37G},
for a recent review see \citealt{2011arXiv1111.0514W}).

Comparatively little work has focused on how the physical motion of
the crust translates into a detectable QPO (although see
\citealt{2008ApJ...680.1398T}). This is not a trivial question: some
of the QPOs have rms fractional amplitudes of up to 20\% of the total
flux in a pulse, whereas the maximum predicted amplitude of a
starquake (assuming the oscillation is restricted to the crust) is
about $0.01 R_*$ (where $R_*$ is the stellar radius). A persistent
starquake will more likely have an amplitude at least 10$\times$
smaller (\citealt{1998ApJ...498L..45D,2011MNRAS.418..659L}; M. Gabler,
2012 private communication).

A solution could come from the steeply peaked pulse profile in the
magnetar tail (see the pulse profile for SGR 1806--20 in
Figure~\ref{fig:ps_rxte}). The observed QPOs are phase dependent, and
are sometimes centered on the steep edge of the pulse (e.g., Figure 3
of \citealt{2005ApJ...632L.111S}). \cite{2005ApJ...632L.111S}
suggested that beamed emission could enhance the observed strength of
an underlying oscillation. As the beam edge sweeps through the
observer's line of sight, the underlying starquake will cause the edge
to wiggle in and out of the observer's line of sight and amplify the
signal.

In this Letter we use a toy model for the emission to quantify the
effect that the pulse shape will have on an underlying small-amplitude
oscillation. The observed oscillation depends strongly both on the
pulse profile and its orientation on the surface of the star. We
demonstrate that the observed oscillation can easily be several times
larger than the underlying physical motion from a starquake. However,
the enhancement is likely not enough to fully explain the observed
amplitudes of the QPOs in SGR 1806--20 and SGR 1900+14, suggesting that
the observed oscillations are produced by a changing intensity in the
emission rather than physical motion of the beam itself.

\section{Model for starquake}
\label{sec:model}

To quantify how the signal from a starquake will be modified by the
pulse profile requires a description for the oscillation and the
emission. We adopt a simplistic model for both the emission and
oscillation geometry, and further assume that the emission comes from
close to the star's surface and moves with the crust. This can easily
be generalized to a more detailed two-dimensional model for both the
crust motion and the pulse pattern.

\begin{figure}
\resizebox{!}{60mm}{\includegraphics[width=80mm]{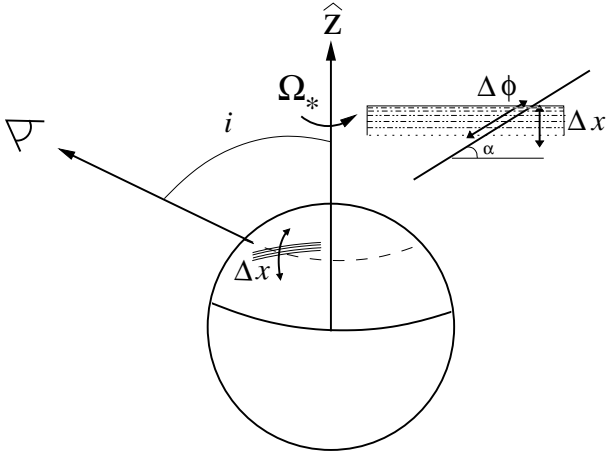}}
\caption{Sketch of star with starquake. Main: the star rotates
  at $\Omega_*$, with a starquake with amplitude $\Delta x$
  superposed on the rotation. The dashed line shows the observer's
  viewing parallel (at inclination $i$). The beam is observed
  (parallel lines on the star's surface) where it crosses the viewing
  parallel. Inset: the viewing parallel (solid line) crosses the
  beam at angle $\alpha$. A physical motion $\Delta x$ is observed as
  a change in phase $\Delta \phi$, which can be large if $\alpha$ is
  small.}
\label{fig:star}
\end{figure}

Figure \ref{fig:star} shows a sketch of the star. The star rotates
with frequency $\Omega_*$ about the $z$-axis, which is inclined by
angle $i$ from the observer's line of sight. The dashed line shows the
observer's viewing parallel: rays parallel to the surface normal on
the viewing parallel will be observable. Assuming the emission is
strongly beamed (necessary to produce sharp gradients), the emission
from the star will be visible where it crosses the viewing parallel.
We assume that the emission beam is a narrow strip in which the
intensity varies strongly across the strip but stays roughly constant
along it. The beam is seen in close-up in the figure and the line
density shows the changing emission intensity.

An oscillation with frequency $\nu_0$ (where $\nu_0\gg\Omega_*$) and
physical displacement $\Delta x$ is superposed on top of the star's
overall rotation. The orientation can be random however we consider
only the component perpendicular to the beam's gradient, since an
oscillation along the gradient will not change the emission
profile. The beam crosses the viewing parallel with angle $\alpha$, so
the observed pulse profile is somewhat broadened.

The high-frequency oscillation $\Delta x$ corresponds to a shift
in the rotation phase, $\Delta \Phi$:
\begin{equation}
\label{eq:delPhi}
\Delta \Phi=\frac{\Delta x}{R_*\sin i\sin\alpha}
\sin(2\pi\nu_0t).
\end{equation}

The $1/\sin\alpha$ term arises from the misalignment between the beam
and viewing parallel. As the beam oscillates in direction $\Delta x$,
the change in flux corresponds to a shift in rotation phase $\Delta
\phi$ larger than the physical displacement. This is because the
observed beam is broadened along the viewing parallel, while $\Delta
x$ probes the `true' beam profile\footnote{We are grateful to Yuri
  Levin for pointing this additional geometrical factor out to
  us.}. Angles $\alpha$ and $\it{i}$ are unknown: however, we will
argue later that they are insufficient to amplify a starquake into the
observed features.


Using the emission profile, $P(\Phi_0)$, we can calculate $P(\Phi)$
and see how the starquake changes the overall variability in the light
curve. Since $\Delta \Phi$ is small, observed profile
is:\begin{eqnarray}
  P(\Phi) &=& P(\Phi_0+\Delta\Phi)\label{eq:qpo_intensity}\\
  \nonumber&=&P(\Phi_0)+\Delta \Phi\left.\frac{dP}{
      d\Phi}\right|_{\Phi_0}+\mathcal{O}(\Delta\Phi^2).
\end{eqnarray}
When the gradient of the light curve is large
($\sim P(\Phi_0)/\langle\Delta\Phi\rangle$), the starquake will be
boosted and dominate the light curve.

This process will have two other effects on the observed
variability. First, since $\frac{dP}{d\Phi}$ changes, the observed
amplitude of the oscillation varies, which will redistribute the
signal power at $\nu_0$ into a range of frequencies. The width of the
signal in the power spectrum can be estimated from the rate of change
in the observed oscillation amplitude, i.e., $\frac{d^2P}{d
  ^2\Phi}$. This process will also introduce harmonics of $\nu_0$ into
the light curve.

In summary: the strength of the oscillation seen by the viewer depends
on $\Delta x$, $i$, $\alpha$ and
$\frac{{\rm~d}P}{{\rm~d}\Phi}\big{|_{\Phi_0}}$. $\frac{d^2P}{d\Phi^2}\big{|_{\Phi_0}}$
determines the width and shape of the feature in the power spectrum,
as well as the magnitude of harmonics of $\nu_0$.

\subsection{Making a Periodic Signal Quasi-Periodic}
\label{sec:qpo}

The high-frequency signals detected in SGR 1806--20 and SGR 1900+14
had resolved widths suggesting that the underlying oscillation is not
a purely periodic signal. To understand how the pulse profile changes
a signal with a range of frequencies, we add QPOs to the light curve
of SGR 1806--20.

We model QPOs as noise processes with a Lorentzian distribution of
powers as input following the method suggested by
\cite{1995A&A...300..707T}. The power spectrum of the QPO in rms
normalization (e.g., \citealt{1989tns..conf...27V}) is given by:
\begin{equation}
\label{eq:qpo}
S_{\rm QPO}=\frac{\sigma^2\Delta~f_{\rm Ny}}{2\pi}\frac{1}{(\Delta/2)^2+(\nu-\nu_0)^2}.
\end{equation}
In Equation~(\ref{eq:qpo}), $f_{\rm Ny}$ is the Nyquist frequency, $\sigma^2$
is the QPO variance, $\Delta$ is its full-width at half-maximum (FWHM),
and $\nu_0$ is its central frequency.

The total power of the QPO is the same as for a sinusoid, only
distributed over a range of frequencies. The variance
of $F(t)=A_0\sin(2\pi\nu_0t)$ over time interval $T$ is:
\begin{equation}
\sigma^2=\frac{A_0^2}{T}\int^{T}_{0} dt \sin^2(2\pi\nu_0 t)=\frac{A_0^2}{2},
\end{equation}
so that substituting the amplitude from Equation~(\ref{eq:delPhi}),
$\sigma^2$ in Equation~(\ref{eq:qpo}) is given by:
\begin{equation}
  \sigma^2=\frac{1}{2}\left(\frac{\Delta x}{R_*}\frac{1}{\sin i\sin\alpha}\right)^2.
\end{equation}

To produce a time series of length $T$, we generate a realization of
its Fourier transform by taking the power (Equation~(\ref{eq:qpo})) at each
frequency, $\nu_n\equiv~n\Delta\nu$ ($\Delta\nu\equiv~1/T$) and
multiplying by a complex number ($A_n+iB_n$) whose real and imaginary
parts are drawn from a Gaussian random distribution with a variance of
one and mean of zero:
\begin{equation}
f_n=(A_n+iB_n)\sqrt{\frac{1}{2}S_{\rm QPO}(\nu_n)}.
\end{equation}
The resulting time series has a variance $\sigma^2$ and a mean of
zero. This signal is the offset $\Delta\Phi$ to the star's phase
$\Phi_0$ as a function of time.

\subsection{Example: A Power Law Light Curve}
\label{sec:test}

We illustrate how the pulse profile changes an underlying oscillation
with an emissionprofile $P(\Phi_0)\propto\Phi^\beta_0$, adopting
different values of $\beta$. This idealized shape corresponds roughly
to the edges of the pulse where QPOs have been observed.

We use the same total flux for each profile, and neglect the effect of
shot noise. We adopt a period of $10~\rm{s}$ for the emission duration
a time resolution of $\Delta t=9.5\times 10^{-6}~{\rm s}$. We add a
sinusoidal oscillation to the rotation phase, with amplitude $\Delta
x/R_*\sin~ i\sin\alpha=0.1$ and frequency $\nu_0=600~{\rm Hz}$.

\begin{figure}
\rotatebox{90}{\resizebox{!}{84mm}{\includegraphics[width=84mm]{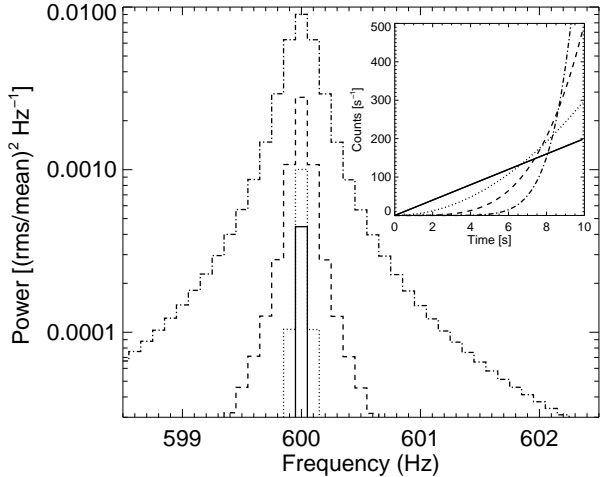}}}
\caption{Main: power spectra centered around high-frequency
  feature for (bottom to top) $\beta=[1,2,4,8]$. The gradient in the
  light curve determines the width and amplitude of the resulting
  feature. Inset: the emission profile for each power
  spectrum. Solid:$\beta=1$; dotted:$\beta=2$;
  dashed:$\beta=4$; dot-dashed:$\beta=8$.\label{fig:ps_ex}}
\end{figure}

Figure \ref{fig:ps_ex} shows the power spectra around $600~\rm{Hz}$,
for $\beta = [1,2,4,8]$, increasing from bottom to top. The inset of
Figure \ref{fig:ps_ex} shows the different emission profiles. The rms
variability from the starquake is measured by integrating the spectrum
in an interval of $30~\rm{Hz}$ centered at $600~\rm{Hz}$ and
subtracting the power contribution from the underlying emission
profile.

The main effect of steeper profiles is to increase the rms variability
of the starquake. For $\beta=1$, the oscillation is a spike at
$600~{\rm~Hz}$ with an rms amplitude of $\sim2\%$. As $\beta$
increases, the amplitude and width of the feature increase, so that
for $\beta=8$ the feature has an rms amplitude of $\sim20\%$---a
ten-fold increase.

For larger $\beta$, the gradient of $P(\Phi_0)$ also changes
substantially, which broadens the feature so that the FWHM for
$\beta=8$ is $0.3~{\rm~Hz}$. The full properties of the signal for
each $\beta$ are summarized in Table \ref{tab:example}.

\begin{table}
\caption{Properties of Oscillation in Power-law Light Curve\label{tab:example}
}
\begin{center}
\begin{tabular}{ l c c c c c}
  \hline
  $\beta$ &$\nu_0$ & Width & rms
  & rms & \\
   & (Hz) & (Hz) & Fundamental& 1st harmonic\\
  \hline
  \hline
  1&600&. . . &2.2\%&0\\
  2&600&0.11&2.9\%&$2.7\times10^{-4}\%$ \\
  4&600&0.162&8.4\%&$1.2\times10^{-3}\%$\\
  8&600&0.3&20.5\%&$6.1\times10^{-3}\%$\\
  \hline
\end{tabular}
\end{center}
\end{table}

From the generated power spectrum we also measure the starquake
harmonics. The last column of Table \ref{tab:example} shows the rms of
the first harmonic (at $1200~\rm{Hz}$) for different $\beta$. Even for
$\beta=8$, the power in the first harmonic is much smaller than the
fundamental and would not be detectable above photon noise.

These results show that a small oscillation can be substantially
increased by a steep gradient of the light curve.We next apply the
same technique to the light curve of SGR 1806--20, to demonstrate how
the beam shape will affect the oscillation amplitude.

\section{The tail of SGR 1806--20}
\label{sec:rxte}

The tail of the 2004 giant flare of SGR 1806--20 showed strongly pulsed
emission with two main peaks in the pulse profile. QPOs appeared at
different phases in the pulse profile, and also lasted for different
lengths of time (\citealt{2006ApJ...637L.117W}, SW06). Two of the
strongest QPOs, one at $\sim90~{\rm~Hz}$ and one at $\sim{\rm
  ~625~Hz}$ are associated with steep gradients in the pulse
profile. A dynamic power spectrum of the data showed that the
$625~{\rm Hz}$ QPO is strongest in the declining edge of the
pulse (Figure 3 of SW06). We therefore focus our study on the same
phase as in SW06.


SW06 found phase-dependent QPOs at $93~{\rm~Hz}$ and $625~{\rm~Hz}$ in
the SGR 1806--20 giant flare beginning $195~{\rm s}$ after the main
spike. The inset of Figure \ref{fig:ps_rxte} shows the averaged pulse
profile for the nine subsequent rotation cycles, and the solid lines
show the phase segment when the QPOs were detected most strongly.

A starquake will only be boosted by rotation-dependent variation of
the light curve. The $RXTE$ data of SGR 1806--20 shows substantial
variability, particularly below 100~Hz, but the rotation period of the
star is closer to $0.1~{\rm Hz}$. Additionally, at shorter time
bins the signal-to-noise ratio drops and the variability
from photon noise dominates. We thus bin and smooth the entire light
curve to a time resolution of 0.1~s, removing signal power at
frequencies above $\sim10~{\rm Hz}$.

Since the resulting feature depends on the gradient of the pulse
profile and its derivative (Equation~(\ref{eq:qpo_intensity})), the binsize
and smoothing change the detailed shape and intensity of the
feature. Experimenting with different binsizes and smoothing
prescriptions shows the amplitude of the feature is robust (to
$\sim10--20\%$) to changes. The width and shape of the feature are
more sensitive to higher derivatives so that the widths and
the shapes of the features we produce serve more as qualitative
examples than quantitative results.

\begin{figure}
\rotatebox{90}{\resizebox{!}{84mm}{\includegraphics[width=84mm]{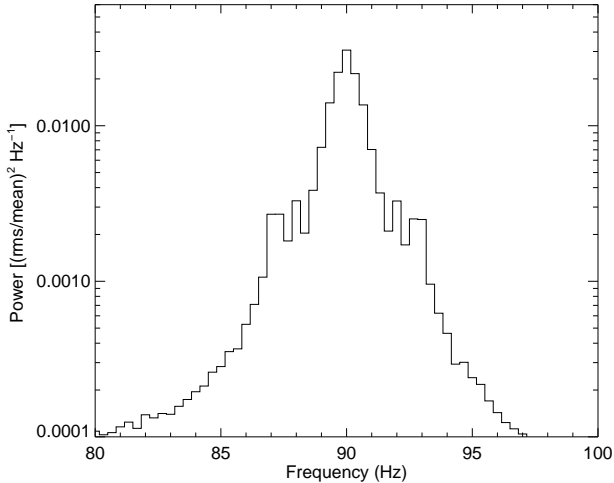}}}
\caption{Feature produced by a sinusoidal oscillation with amplitude
  0.1 and frequency 90~Hz added to the phase for the light curve
  profile of SGR 1806--20.}
\label{fig:sgr_sine}
\end{figure}

To probe the effect of the pulse shape on the ouput oscillation
strength, we make a light curve with two features, with
$\Delta\Phi=0.1\sin(2\pi\nu_0~t)$ at $\nu_0=90~{\rm~Hz}$ and at
$\nu_0=625~{\rm~Hz}$. We then follow SW06 and examine the light curve
$195-260~\rm{s}$ after the main flare, calculating the power spectrum
of the 3~s phase segment shown in the inset of
Figure~\ref{fig:ps_rxte}. The rms amplitude of the signal changes
between rotations (from the changing pulse profile) from
$16\%$ to $25\%$. Stacking and averaging the power spectra we measure the
average power in each feature. The stacked power spectrum centered
around $90~\rm{Hz}$ is shown in Figure \ref{fig:sgr_sine}. The feature
has an amplitude of $21\%$ and FWHM width of $1.6~\rm{Hz}$. The
feature at $625~\rm{Hz}$ has an rms amplitude of $20\%$ and width
$1.6~\rm{Hz}$. While the width is still smaller than the widths of the
measured features ($1.8~{\rm Hz}$ for the $625~\rm{Hz}$ feature and
$8.8~{\rm Hz}$ for the $93~\rm{Hz}$ feature; SW06), it nonetheless
demonstrates that a very narrow feature can be substantially broadened
by the light curve profile.

The harmonics of the features are stronger than for the power-law case
with an rms amplitude of $3\%$ and a large frequency spread in the
first harmonic. This is still undetectable, even when the oscillation
has an observed rms amplitude of $~20\%$.

\subsection{QPOs in SGR 1806--20}
\label{sec:result}

The QPO features detected in SGR 1806--20 have measurable widths
between $1--17~\rm{Hz}$, and frequencies where photon noise dominates
the variability. Both these effects can change the measured properties
of QPO, as we study below. We model the starquake as a quasi-periodic
signal as outlined in Section~\ref{sec:qpo}, defining an input width,
central frequency and amplitude. To simulate photon noise we use the
output light curve as a probability function to detect a photon in a
given time bin and generate a light curve as a series of discrete
events. Since these are both random noise processes, we generate a
large number of realizations and measure the QPO properties
statistically.

As above, we add two QPOs to the phase of the SGR 1806--20 light curve
and calculate the modified profile. One QPO is centered at
$93~\rm{Hz}$ with a width of $6~{\rm Hz}$, the second at $625~\rm{Hz}$
with a width of $2~{\rm Hz}$. Both have an input amplitude of
$|\Delta\Phi|= 0.1$. We use the same rotation phase as above, stack
the power spectra from nine successive rotations and bin the data with
a binsize of $2.66~\rm{Hz}$.

\begin{figure}
\rotatebox{90}{\resizebox{!}{84mm}{\includegraphics[width=84mm]{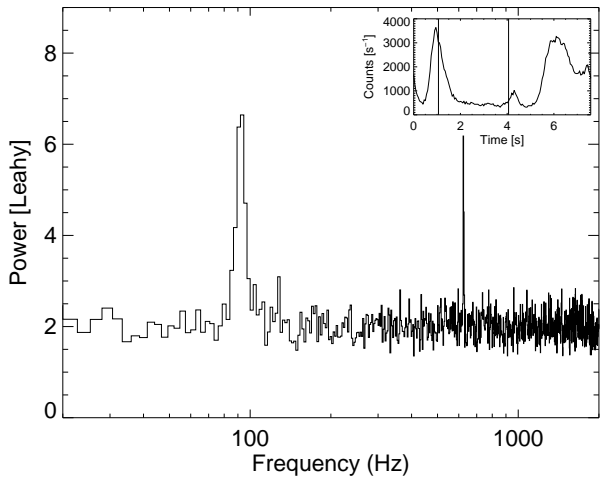}}}
\caption{Main: power spectrum of a simulation using the SGR
  1806-20 light curve (smoothed to a resolution of 0.1s). QPOs were
  inserted at 93~Hz and 626~Hz. Inset: the rotation-averaged
  pulse profile 195-260~s after the main flare. The section of
  the phase profile used in the power spectrum is marked by vertical
  lines.\label{fig:ps_rxte}}
\end{figure}

Figure~\ref{fig:ps_rxte} shows the power spectrum (in Leahy
normalization) of one realization. The inset shows the averaged phase
profile , with vertical lines denoting the phase segment used to make
the power spectrum. This figure is comparable to Figure~2 of
SW06. Both QPOs are visible by eye. In our spectrum the power above
$\sim10~\rm{Hz}$ has been removed from the smoothing applied to the
light curve.

The QPO inserted at 93~Hz is detected with a mean width of
6.6$\pm$1.5~Hz, rms amplitude of 21\%$\pm$1.5\%, and average central
frequency of 91.7$\pm$0.6~Hz. For the 626~Hz QPO, the mean width is
2.4$\pm$1.4~Hz, with an rms amplitude of 15\%$\pm$1.5\% and central
frequency 624.9$\pm$0.35~Hz.

The rms amplitude at 93~Hz is not significantly different from adding
a sinusoid, but is measurably lower around 626~Hz. In both cases,
however, the strength of the feature is significantly stronger than
the underlying oscillation. The features show modest broadening with a
large dispersion between different realizations. The systematic offset
in the central frequency of the QPO appears because of the binsize. In
simulations without photon noise, reducing the binsize shifts the
central frequency of the QPO closer to the input one. Harmonics are
not detected.

\section{Discussion}
\label{sec:discussion}

The steep pulse profile seen in the tail of giant magnetar flares
substantially alters the observed properties of a starquake, changing
both its observed amplitude and power distribution. While this effect
is not large enough to transform a starquake with maximum amplitude
$\Delta~x/R_*\sim0.001-0.01$ into a QPO with an rms amplitude of 20\%, the
boosting can still be considerable, especially if the emission comes
from close to the rotation axis of the star, or if the steepest part
of the beam gradient is misaligned with the viewing parallel (small
$\alpha$).

There are several uncertainties in our calculation that could boost
the amplitude of a starquake further. The first of these is the
unknown location of emission on the surface of the star. If the
emission is close to the rotation axis, then a small physical motion
produces a larger change in phase than at the equator. However, to
produce a ten-fold increase requires an inclination angle of less than
$\sim6^\circ$, which is improbable given the strong observed
pulsations.

The second way to boost the amplitude is for the beam to cross the
observing parallel obliquely, as in Figure~\ref{fig:star}. Since in
this case the actual beam edge is steeper than observed, the
amplification is larger than expected from the observed beam
shape. For the extreme case ($\alpha=0$) the beam runs exactly
parallel to the line of sight so that no pulse is seen but there is a
strong QPO. This scenario is unlikely, both that $\alpha$ is so small
and that the beam edge is so much sharper (by a factor $1/|\sin
\alpha|$) than observed. As an example: to produce a feature with a
rms amplitude of $20\%$ requires a gradient
$0.2\sim dP/d\Phi\sim\langle~P_0\rangle~A_0$ (cf. Equation
(\ref{eq:qpo_intensity})), which is $\sim100\times$ steeper than the
pulse profile in Figure~\ref{fig:ps_rxte}.

The input amplitudes required to produce the strong QPOs of SGR
1806-20 are large enough to produce QPOs that are observable
throughout the entire profile. Applying the model across the entire
pulse profile (using 3s segments offset by $0.1P_*$ in phase), the oscillation
has an output amplitude between $10\%$ and $25\%$--strong enough to be
detected throughout the rotation. If the emission model is correct,
this requires that the starquake amplitude varies significantly across
the surface of the crust and the correlation with the falling pulse
edge is somewhat coincidental. This is expected generically in
starquake models, however, and is seen directly in MHD simulations of
coupled core-crust oscillations \citep{2011MNRAS.410L..37G} which show
a strong phase dependence in the starquake amplitude.

The toy model suggests that the pulse profile can substantially
influence the observed amplitude and other properties of a
starquake. While we have used a simple model for the emission and
starquake, this work can be easily generalized to use a
physically-motivated model of the surface emission and starquake, in
order to put constraints on the pattern of emission on the surface of
the star during the quake.

We thank Phil Uttley, Yuri Levin and Daniela Huppenkothen for useful
discussion and acknowledge support from an NWO Vidi Grant.


\end{document}